\documentstyle[12pt,epsf,times]{article}

\renewcommand{\l}{\lambda}
\renewcommand{\k}{\kappa}

\begin{document}
\begin{titlepage}
\def\thefootnote{\fnsymbol{footnote}}

\rightline{ITP-UH-07/95}
\rightline{February 1995}

\vspace*{\fill}
\begin{center}
{\Large Integrable models of coupled Heisenberg chains} \\
\vfill
\vspace{1.5 em}
{\sc Holger Frahm}\footnote{e-mail: {\tt frahm@itp.uni-hannover.de}}
and 
{\sc Claus R\"odenbeck}\footnote{e-mail: {\tt roeden@itp.uni-hannover.de}}\\ 
{\sl Institut f\"ur Theoretische Physik, Universit\"at Hannover\\
D-30167~Hannover, Germany}\\
\vfill
ABSTRACT
\end{center}
\begin{quote}
\setlength{\baselineskip}{13pt}
We show that the solutions of the Yang--Baxter equation invariant under the
action of the Yangian $Y(sl_2)$ lead to inhomogenous vertex models.
Starting from a four dimensional representation of $Y(sl_2)$ we obtain an
integrable family of coupled Heisenberg spin-$1\over2$ chains. Some
thermodynamical properties of this model are studied by means of the
algebraic Bethe Ansatz.
\end{quote}

\vfill
PACS-numbers: 
75.10.Jm~\      
05.30.-d~\      
03.65.-w~\      

\vfill
\setcounter{footnote}{0}
\end{titlepage}

Integrable models of one dimensional quantum systems and the related two
dimensional classical statistical models have provided much of the insight
into correlation effects in low dimensional systems. In particular, Bethe
Ansatz methods have been used to study the thermodynamical properties of
these systems as well as the excitation spectrum and asymptotic behaviour
of correlation functions.

The construction of integrable models within the Quantum Inverse Scattering
Method (QISM) is based on solutions of the Quantum Yang Baxter equation
\cite{fadd:80} 
\begin{equation}
   R_{12}(\lambda-\mu)\ R_{13}(\lambda)\ R_{23}(\mu)
	=R_{23}(\mu)\ R_{13}(\lambda)\ R_{12}(\lambda-\mu)\ .
\label{qybe}
\end{equation}
The central objects in this approach are the so called $R$-matrices:
$R_{ij}(\lambda)$ is a linear operator acting on $V_i \otimes
V_j$ depending on a complex {\em spectral parameter} $\lambda$. Hence
(\ref{qybe}) is an equation on the product space $V_1 \otimes V_2\otimes
V_3$ with certain finite dimensional vector spaces $V_i$. Given a solution
of (\ref{qybe}) one can construct a family of commuting operators (which
include the Hamiltonian) and compute the spectrum using the algebraic Bethe
Ansatz.

Particular solutions of Eq.(\ref{qybe}) with rational dependence on the
spectral parameter are found by looking for $R$-matrices invariant under
the action of a simple Lie algebra \cite{kulx:81}. In this case $V_i$ is to
be identified as the representation space of an irreducible finite
dimensional representation of this algebra. For the case of $sl_2$ this
construction leads to the series of higher spin $SU(2)$ Heisenberg chains
\cite{babu:82,takh:82} starting with the $S={1/2}$ Hamiltonian
\begin{equation}
   {\cal H} = \sum_{n}
      \vec{S}_n\cdot\vec{S}_{n+1} -{1\over4}\ .
\label{hheis}
\end{equation}

On the other hand Eq.\ (\ref{qybe}) for given $R_{12}$ can be interpreted
as the definition of a quadratic algebra of operators on $V_3$. The
resulting Hopf algebra is called a Yangian. In the following we shall
concentrate on the Yangian associated to $sl_2$ which will eventually
result in the construction of $SU(2)$-invariant soluble spin chains.

For the simplest so called {\em evaluation representations} $V_m(a)$ of the
Yangian $Y(sl_2)$ the generators $x$ and $J(x)$ of the algebra can be given
in terms of irreducible $m+1$-dimensional representations of $sl_2$ with
generators $\{x^+, x^-, h\}$ and an arbitrary complex number
$\alpha$\footnote{For later convenience we rotate the parameter $\alpha$
as compared to the notation used in \cite{chpr:90}}
\begin{equation}
   x \in \{x^+, x^-, h\}\ , \quad J(x) = i\alpha x\ .
\label{y_eval}
\end{equation}
As was shown by Chari and Pressley every finite dimensional irreducible
representation of $Y(sl_2)$ is isomorphic to a tensor product of these
evaluation representations \cite{chpr:90}.

Associated to such a representation of the Yangian Drinfel'd has proven
that there exists a unique ``universal $R$-matrix'' which is a rational
function of the spectral parameter \cite{drin:85}. Later, Chari and
Pressley wrote down explicitly the solution of Eq. (\ref{qybe}) associated
to a finite-dimensional irreducible representation of $Y(sl_2)$, i.e.\
$R$-matrices intertwining between two copies of this representation.

In this letter we show that these $R$-matrices can be interpreted as
plaquettes containing several vertices of certain inhomogeneous 6 vertex
models. This fact allows for the diagonalization of the corresponding
transfer matrices using the algebraic Bethe Ansatz. Specifically we study
the Hamiltonian (which is integrable by construction) of the quantum spin
chain corresponding to the simplest representation of the Yangian beyond
the ones given in (\ref{y_eval}) (which result in the known spin-${m/2}$
Heisenberg models) namely tensor products of two evaluation representations
$V_1(a)\otimes V_1(b)$. It turns out that the resulting model is a system
of two $S={1/2}$ chains with Hamiltonian (\ref{hheis}) coupled by two and
three spin exchange terms with coupling constants depending on a real
parameter.

The approach used in Ref. \cite{chpr:90} to compute the solutions of
(\ref{qybe}) depends on the concept of the {\em intertwiner}
$I(\alpha-\beta)$: Given a tensor product of two evaluation representations
$V_m(\alpha)$ and $V_n(\beta)$ the intertwiner is uniquely defined through
the following properties: It interchanges the order of the factors, i.e.\
$V_n(\beta) \otimes V_m(\alpha) \longrightarrow V_m(\alpha)\otimes
V_n(\beta)$, the Yangian highest weight state $\Omega_n\otimes \Omega_m$ is
mapped onto $\Omega_m\otimes \Omega_n$ and $I(\alpha-\beta)$ commutes with
all elements of the Yangian.

Considered as a representation of $sl_2$ $V_n\otimes V_m$ is isomorphic to
the direct sum
\begin{equation}
   V_n \otimes V_m\cong\bigoplus_{j=0}^{\mbox{min}\{m,n\}}V_{m+n-2j}.
\end{equation}
By definition the intertwiner commutes with the action of $sl_2$. Therefore
it can be expanded into a sum of projectors on the irreducible components
of the product $V_m\otimes V_n$:
\begin{equation}
   I(\alpha-\beta) = \sum_{j=0}^{\mbox{min}\{m,n\}}c_jP_{m+n-2j},
\end{equation}
where $P_{m+n-2j}:\ V_n \otimes V_m \longrightarrow V_m\otimes V_n$
projects into the spin $(m+n-2j)/2$-component of $V_m\otimes V_n$. The
coefficients $c_j$ are given by the recursion relation \cite{kulx:81}
\begin{equation}
   \frac{c_j}{c_{j-1}}
  =\frac{i\alpha-i\beta+\frac{1}{2}(m+n)-j+1}{i\alpha-i\beta-\frac{1}{2}(m+n)+j-1}.
\end{equation}
The $R$-matrix associated to the product $V_m(\alpha)\otimes V_n(\beta)$ then is given by
\begin{equation}
   R(\lambda)=I(\lambda)\ \sigma,
\label{eq:Reval}
\end{equation}
$\sigma$ being the `switch map'. These $R$-matrices associated to
evaluation representations were first calculated by Kulish {\em et
al.}\cite{kulx:81}. 

As shown in Ref.~\cite{chpr:90} $R$-matrices associated to {\em any}
tensorproduct of evaluation representations and hence {\em any} irreducible
representation of $Y(sl_2)$ are uniquely determined by a product of
$R$-matrices (\ref{eq:Reval}) involving only two factors of the
productspace, respectively.

As an example consider the tensor product of four dimensional representations
$V_A(\lambda)=V_1(\lambda)\otimes V_1(\lambda+\kappa)$ and
$V_B(\mu)=V_1(\mu)\otimes V_1(\mu+\epsilon)$ of $Y(sl_2)$
\begin{equation}
  V_A(\lambda)\otimes V_B(\mu) =
  \overbrace{
    \underbrace{V_1(\lambda)}_1\otimes
    \underbrace{V_1(\lambda+\kappa)}_2}^A\otimes
  \overbrace{
    \underbrace{V_1(\mu)}_3\otimes
    \underbrace{V_1(\mu+\epsilon)}_4}^B\ .
\label{prodspace}
\end{equation} 
Following Chari and Pressley  the $R$-matrix on this space is given by
\begin{equation}
   R_{AB}(\lambda) \equiv R_{23}(\lambda+\kappa)\
                          R_{24}(\lambda+\kappa-\epsilon)\
                          R_{13}(\lambda)\
                          R_{14}(\lambda-\epsilon)
\label{eq:RAB}
\end{equation}
where $R_{kl}$ is an $R$-matrix of type (\ref{eq:Reval}) on the $k$-th and
$l$-th factor in $V_A\otimes V_B$. For the special case of two dimensional
(spin $1\over 2$) evaluation representations $V_1$ it is Yang's R-matrix
\cite{yang:67}
\begin{equation}
   R_{kl}(\lambda)=\frac{\lambda-i\sigma_{kl}}{\lambda-i}\ .
\end{equation}
$\sigma_{kl}$ interchanges space $k$ and $l$.

\begin{figure}[ht]
\begin{center}
\leavevmode
\epsfbox{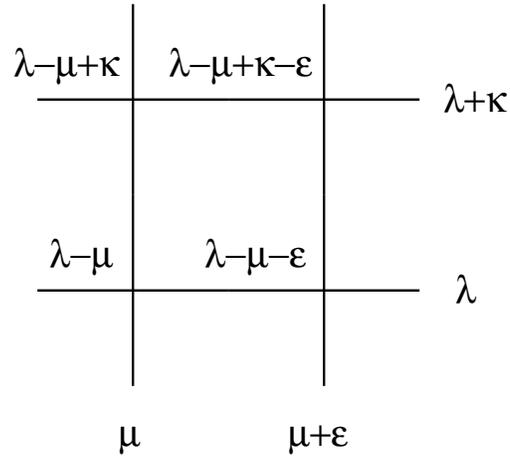}
\end{center}
\caption{\label{fig:plaq}
Plaquette of elementary vertices in the $R$-matrix $R_{AB}(\lambda-\mu)$
(\ref{eq:RAB})}
\end{figure}

In the language of vertex models $R_{kl}(\lambda-\mu)$ can be represented
graphically by
\begin{center}
\leavevmode
\epsfbox{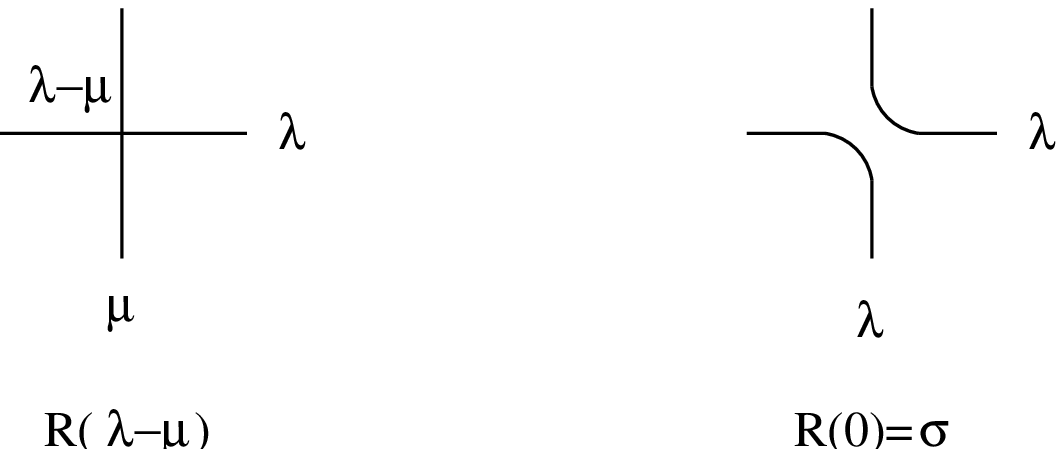}
\end{center}
where one identifies horizontal lines with the {\em matrix space} $k$ and
vertical ones with the {\em quantum space} $l$.  In this framework the
$R$-matrix (\ref{eq:RAB}) is just a group of vertices in an inhomogeneous
vertex model (Fig.~\ref{fig:plaq}).

Multiplying in matrix space $N$ of these objects acting on different copies
of the quantum space gives the monodromy matrix ${\cal T}_{AB}(\lambda-\mu)$.
By construction ${\cal T}_{AB}$ satisfies the Yang-Baxter equation
\begin{equation}
    R_{AB}(\lambda-\mu)\
      ({\cal T}_{AB}(\lambda)\otimes 1)\ (1\otimes{\cal T}_{AB}(\mu))\ =\
    (1\otimes{\cal T}_{AB}(\mu))\ ({\cal T}_{AB}(\lambda)\otimes 1)\
         R_{AB}(\lambda-\mu)\ .
\label{YBE}
\end{equation}
As a direct consequence the transfer matrix $T_{AB}(\lambda)=tr\,{\cal
T}_{AB}(\lambda)$ commutes for different values of the spectral parameter
$\lambda$. Here the trace is taken with respect to the matrix
space. Therefore $T_{AB}(\lambda)$ can be interpreted as a generating
functional of commuting operators in the quantum space.

\begin{figure}[ht]
\begin{center}
\leavevmode
\epsfxsize=\textwidth
\epsfbox{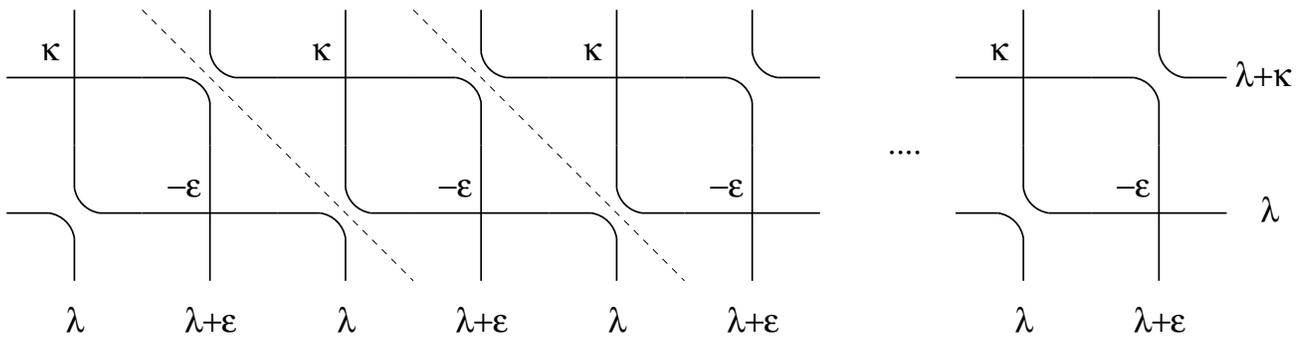}
\end{center}
\caption{\label{fig:mono}
Graphical representation of ${\cal T}_{AB}(\lambda)$.}
\end{figure}

Obviously, ${\cal T}_{AB}$ can be written as (see also Fig.~\ref{fig:mono})
\begin{equation}
   {\cal T}_{AB}(\lambda)=\tau(\lambda+\kappa)\otimes\tau(\lambda)
\label{eq:monoAB}
\end{equation}
where $\tau(\lambda)$ is the monodromy matrix of the inhomogeneous vertex
model constructed from $R_{1B}(\lambda) \equiv R_{13}(\lambda)\
R_{14}(\lambda-\epsilon)$ (see (\ref{prodspace}) for the numbering of the
different spaces).
From (\ref{eq:monoAB}) we obtain the following relation between $T_{AB}$
and the transfer matrix $t(\lambda)=tr\,\tau(\lambda)$
\begin{equation}
   T_{AB}(\lambda)=t(\lambda+\kappa)\,t(\lambda)\ .
\label{therelforT}
\end{equation}
Since $R_{1B}$ itself is a solution of a Yang--Baxter equation
(\ref{qybe}) we have $[t(\lambda),t(\mu)]=0$. Consequently the
diagonalization of $T_{AB}$ is equivalent to the solution of the
eigenvalue problem for $t(\lambda)$.

The latter is obtained in the framework of the algebraic Bethe Ansatz (see
e.g.\ \cite{fata:84}). Starting from the ferromagnetic vacuum
$|\Omega\rangle = |\uparrow\uparrow\ldots\uparrow\rangle$ which is an
eigenstate of $t_{AB}(\lambda)$ with eigenvalue
\begin{equation}
   \widetilde\Lambda_0(\lambda) = 
     1 + \left(\frac{\lambda}{\lambda+i}\right)^N
   \left(\frac{\lambda-\epsilon}{\lambda-\epsilon+i}\right)^N
\end{equation}
one constructs eigenstates of $T_{AB}(\lambda)$ with $M$ overturned spins
parametrized by $M$ rapidities $\lambda_j$ with eigenvalues
\begin{equation}
   \Lambda(\lambda)
       =\widetilde{\Lambda}(\lambda)\widetilde{\Lambda}(\lambda+\kappa)\ .
\label{eigenvalue}
\end{equation}
Here $\tilde\Lambda$ are the eigenvalues of $t(\lambda)$
\begin{equation}
   \widetilde{\Lambda}(\lambda)=\prod_{j=1}^l
   \frac{\lambda_j-\lambda+\frac{i}{2}}{\lambda_j-\lambda-\frac{i}{2}}+
   \left(\frac{\lambda}{\lambda+i}\right)^N
   \left(\frac{\lambda-\epsilon}{\lambda-\epsilon+i}\right)^N
   \prod_{j=1}^l
    \frac{\lambda-\lambda_j+\frac{3}{2}i}{\lambda-\lambda_j+\frac{i}{2}}
\end{equation}
and the $\lambda_j$ are solutions of the Bethe Ansatz equations
\begin{equation}
   \left(\frac{\lambda_j+\frac{i}{2}}{\lambda_j-\frac{i}{2}}\right)^N
   \left(\frac{\lambda_j-\epsilon+\frac{i}{2}}{\lambda_j-\epsilon-\frac{i}{2}}\right)^N
   =\prod_{j\not= k}\frac{\lambda_j-\lambda_k+i}{\lambda_j-\lambda_k-i}\ .
\label{BAEQ}
\end{equation}

For the construction of a Hamiltonian with local interactions it is
necessary that $T_{AB}(\lambda)$ has a {\em shift point}, i.e.\ a value of
the spectral parameter $\lambda$, where it degenerates to the shift
operator. As shown in Fig.~\ref{fig:mono} for $\lambda=0$ configurations
in quantum space are shifted with an additional coupling between pairs of
neighbouring spins. Hence the necessary condition for the existence of a
shift point is that this coupling becomes the identity.

As a direct consequence of the Yang-Baxter equation (\ref{qybe}) the
$R$-matrices (in the normalization used here) satisfy the unitarity
condition
\begin{equation}
   R(\lambda)\,\sigma R(-\lambda)\,\sigma=1\ .
   \label{eq:uniR}
\end{equation}
Consequently, choosing $\epsilon=\kappa$ we obtain a one-parameter family of
monodromy matrices allowing for the construction of Hamiltonians with
local interactions.

\begin{figure}[ht]
\begin{center}
\leavevmode
\epsfxsize=0.8\textwidth
\epsfbox{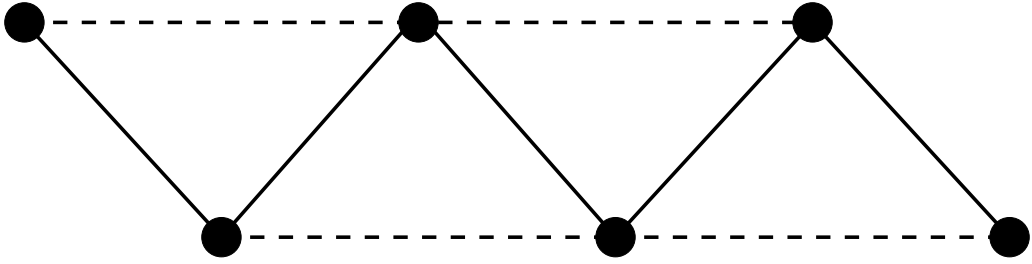}
\end{center}
\caption{\label{fig:lattice}
Lattice on which the spin Hamiltonian (\ref{eq:Hsc}) is defined: The
two-spin exchange coupling is $2J_1$ and $J_3$ on full and dashed lines,
respectively. The three spin exchange $\propto J_2$  couples the spins on
the corners of each triangle.}
\end{figure}

The Hamiltonian of the integrable spin chain is obtained by taking the
logarithmic derivative of the transfer matrix $T_{AB}(\lambda)$ at the
shiftpoint $\lambda=0$:
\begin{equation}
   \frac{i}{2}\frac{d}{d\lambda}\ln T_{AB}(\lambda)\vert_{\lambda=0}
       	\equiv {\cal H}(\kappa)\ .
\end{equation}
The resulting spin chain Hamiltonian is ${\cal H}(\kappa)\equiv\sum_{n=1}^N
{\cal H}_n(\kappa)$ with
\begin{eqnarray}
   {\cal H}_n(\kappa)&\equiv&
   J_1\left(\vec{S}_{2n-1}\vec{S}_{2n}+2\vec{S}_{2n}\vec{S}_{2n+1}
    +\vec{S}_{2n+1}\vec{S}_{2n+2}-1\right)+\nonumber \\
  &&+J_2\left(\vec{S}_{2n}(\vec{S}_{2n+2}\times\vec{S}_{2n+1})+\vec{S}_{2n-1}
   (\vec{S}_{2n}\times\vec{S}_{2n+1})\right)+\nonumber \\
   &&+J_3\left(\vec{S}_{2n}\vec{S}_{2n+2}+\vec{S}_{2n-1}\vec{S}_{2n+1}
	-{1\over2}\right)\ .
\label{eq:Hsc}
\end{eqnarray}
The exchange couplings $J_i$ depend on the parameter $\kappa$ in the
following way
\begin{equation}
   J_1=\frac{1}{1+\kappa^2}\ , \quad 
   J_2=\frac{2\kappa}{1+\kappa^2}\ , \quad 
   J_3=\frac{\kappa^2}{1+\kappa^2}\ .
\end{equation}
For $\kappa=0$ ${\cal H}$ reduces to the Hamiltonian (\ref{hheis}) of a
spin ${1\over2}$ Heisenberg model on a chain of length $L=2N$, for
$\kappa\to\infty$ one obtains two decoupled systems of length $L=N$
each. For intermediate values of the coupling $\kappa$ one has a system of
two coupled Heisenberg chains (see Fig.~\ref{fig:lattice}).

From (\ref{eigenvalue}) the eigenvalues of (\ref{eq:Hsc}) are obtained as
\begin{equation}
   E(\{\lambda_j\}) =
  \frac{i}{2}{\partial\over\partial\lambda}\ln\Lambda(\lambda)|_{\lambda=0}
  = \sum_j \left(\tilde{\epsilon}^{(0)}(\l_j)
	        +\tilde{\epsilon}^{(0)}(\l_j-\k)\right)\ ,\quad
   \tilde{\epsilon}^{(0)}(\l_j)=-\frac{1}{2}\ \frac{1}{\l_j^2+\frac{1}{4}} \ .
  \label{eigenvH}
\end{equation}

In the thermodynamic limit the solutions of the Bethe Ansatz equations
(\ref{BAEQ}) are known to be arranged in bound states if uniformly spaced
sets of complex $\lambda_j$, so called {\em strings}:
\begin{equation}
   \lambda_j^{(m)}=x+i\mu_j\ ,
   \qquad\mu_j=\frac{-m+1}{2},\frac{-m+3}{2},\ldots,\frac{m-1}{2}.
\end{equation}
As in the Heisenberg model the ground state (in absence of a magnetic
field) is made up of $N$ real $\lambda$'s only. Their densities
$\rho(\lambda)$ can be given in terms of a linear integral equation
resulting in
\begin{equation}
   \rho(\lambda) = \tilde{\rho}(\lambda)+\tilde{\rho}(\lambda-\kappa)\ ,\quad
   \hbox{with }\tilde{\rho}=\frac{1}{2\cosh(\pi\lambda)}\ .
\end{equation} 
From (\ref{eigenvH}) the ground state energy $E_0$ per spin is given by
\begin{eqnarray}
   {E_0\over {2N}}&=&
     {1\over2}\int_{-\infty}^{+\infty}d\lambda\,
           \rho(\lambda)\epsilon^{(0)}(\lambda)
  = \int_{-\infty}^{+\infty}d\lambda\,
          (\tilde{\rho}(\lambda)\tilde{\epsilon}^{(0)}(\lambda)
          +\tilde{\rho}(\lambda-\kappa)\tilde{\epsilon}^{(0)}(\lambda))
  \nonumber \\
  &=& \ln 2 + {1\over2}
	\left(\beta(1-i\kappa)+\beta(1+i\kappa)\right)\ .
\end{eqnarray}
The term containing $\beta$-functions
\begin{equation}
   \beta(x)=\frac{1}{2}\left(\psi\left(\frac{x+1}{2}\right)
                            -\psi\left(\frac{x}{2}\right)\right)
\end{equation}
($\psi(x)$ being the digamma function) decreases from $\ln 2$ to 0 as
$\kappa$ varies between 0 and $\infty$.

\begin{figure}[ht]
\begin{center}
\leavevmode
\epsfxsize=0.7\textwidth
\epsfbox{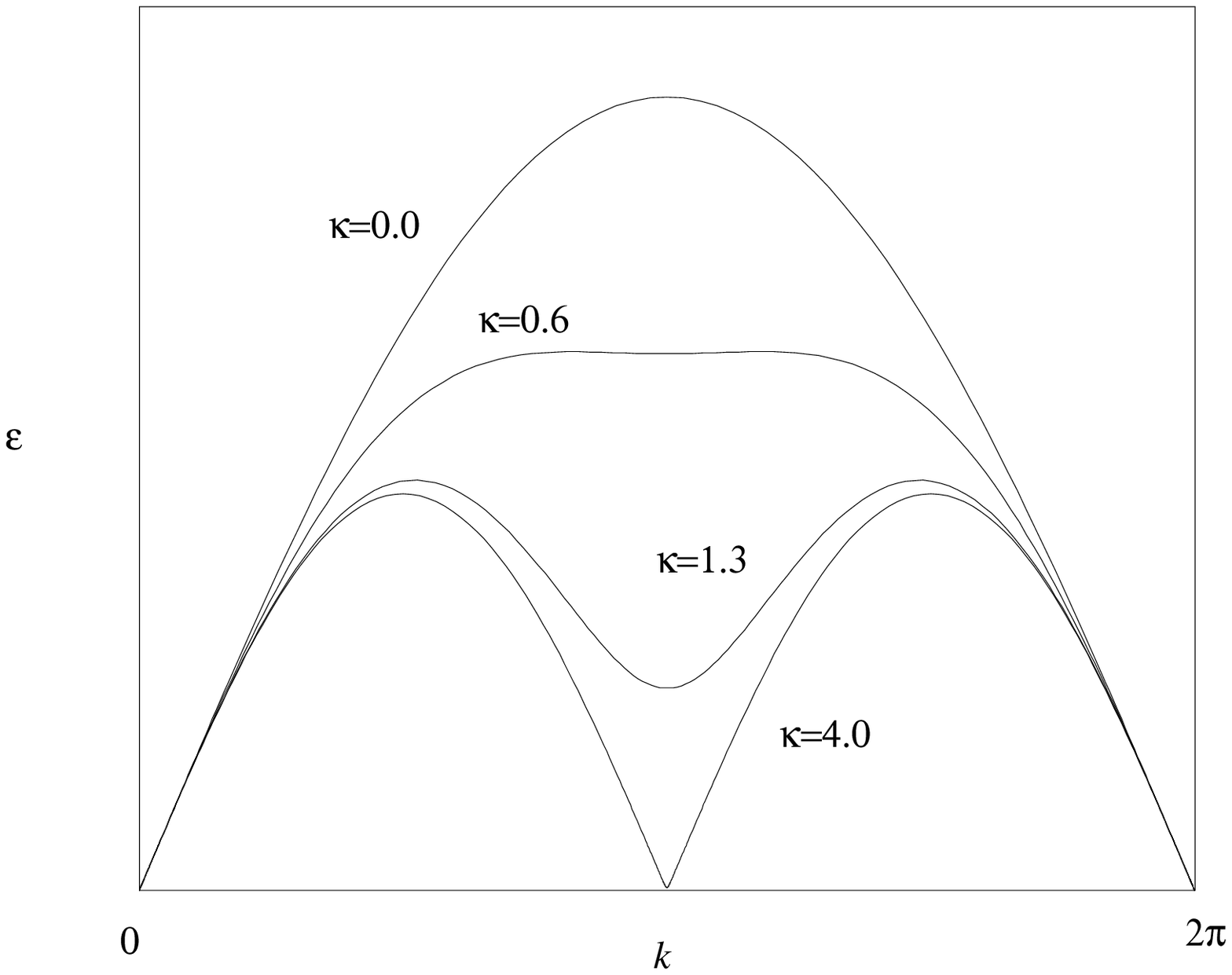}
\end{center}
\caption{\label{fig:disp}
Spinon dispersion for several values of $\kappa$}
\end{figure}

As in the Heisenberg model the low lying excitations above the ground state
are made up of an even number of spin-$1/2$ objects \cite{fata:81}
called {\em spinons}. These elementary excitations are holes in the
distribution of 1-strings parametrized by a rapidity $\lambda$. Their
dispersion is found to be
\begin{eqnarray}  
   \epsilon(\lambda)=\tilde\epsilon(\lambda)+\tilde\epsilon(\lambda-\k)\ ,
   &\qquad&
   \tilde\epsilon(\l)=\frac{\pi}{2\cosh(\l)} \ \nonumber \\
   k(\lambda) = \tilde{k}(\lambda) + \tilde{k}(\lambda-\kappa)\ ,
   &\qquad&
   \tilde{k}(\l)=\arctan(\l)-\frac{\pi}{2} \ .
\label{spinondisp}
\end{eqnarray}
In Fig.~\ref{fig:disp} the spinon dispersion $\epsilon(k)$ is shown for
several values of the parameter $\kappa$.

Several extensions to the system discussed in this paper are possible:

First, the above methods are easily extended to representations
$V_m(a)\otimes V_n(b)$ of $Y(sl_2)$. This leads to one parametric families
of integrable quantum spin chains with alternating spin $S={m/2}$ and
$S={n/2}$. In fact, the case of $m=1$, $n=2$ is the isotropic limit
of the alternating spin model constructed by de Vega and
Woynarovich \cite{vewo:92} at its conformal point\footnote{%
In Ref.~\cite{vewo:92} two commuting Hamiltonian operators $\bar{H}$ and
$\tilde{H}$ are constructed for each value of $\kappa$. The method used
here produces the rotational invariant vertex model with ${\cal H}
=\bar{H}+\tilde{H}$.}
(for further studies of this model and its generalization to $m>2$ at
coupling corresponding to $\kappa=0$ see also
Refs.~\cite{alma:93,vegx:94}). In that work the factors in (\ref{eq:RAB})
are chosen to be the $R$-matrices arising in the context of the higher spin
XXZ models. Again unitarity relation (\ref{eq:uniR}) is used to construct
families of commuting transfer matrices (see also \cite{anjo:84}).  While
the Yangian approach is limited to $R$-matrices with rational dependence on
the spectral parameter an extension to hyperbolic $R$-matrices using the
approach of Refs.~\cite{anjo:84,vewo:92} is straightforward.

Second, irreducible representations of $Y(sl_2)$ of the type
$V_{m_1}(a_1)\otimes\ldots\otimes V_{m_k}(a_k)$ built from more than two
evaluation representations (\ref{y_eval}) can be considered to construct
integrable models of $k$ coupled spin chains. While additional multi--spin
coupling terms are likely to arise within this construction we expect $k-1$
free parameters (similar to $\kappa$) which can be adjusted to remove some
of these interaction terms from the Hamiltonian.

Finally, the construction of integrable Hamiltonians with local
interactions from inhomogeneous vertex models such as the one given in
Fig.~\ref{fig:plaq} allows for the generalization to exchange symmetries
other than $sl_2$ including supersymmetric extensions leading to $t$-$J$
and extended Hubbard models on coupled chains.

This work has been supported in part by the Deutsche
Forschungsgemeinschaft under Grant No.\ Fr~737/2--1.


\newpage
\section*{Figures}
\vfill

\vfill

\vfill

\begin{thebibliography}{10}

\bibitem{fadd:80}
L. Faddeev, Sov. Sci. Rev. C {\bf 1},  107  (1980).

\bibitem{kulx:81}
P.~P. Kulish, N.~Yu. Reshetikhin, and E.~K. Sklyanin, Lett. Math. Phys. {\bf 5},
   393  (1981).

\bibitem{babu:82}
J. Babujian, Phys. Lett. A {\bf 90},  479  (1982).

\bibitem{takh:82}
L. Takhtajan, Phys. Lett. A {\bf 87},  479  (1982).

\bibitem{chpr:90}
V. Chari and A. Pressley, L'Enseignement Math. {\bf 36},  267  (1990).

\bibitem{drin:85}
V.~G. Drinfel'd, Soviet Math. Dokl. {\bf 32},  254  (1985).

\bibitem{yang:67}
C.~N. Yang, Phys. Rev. Lett. {\bf 19},  1312  (1967).

\bibitem{fata:84}
L.~D. Faddeev and L.~A. Takhtajan, J. Sov. Math. {\bf 24},  241  (1984), [Zap.
  Nauch. Semin. LOMI {\bf 109}, 134 (1981)].

\bibitem{fata:81}
L.~D. Faddeev and L.~A. Takhtajan, Phys. Lett. A {\bf 85},  375  (1981).

\bibitem{vewo:92}
H.~J. de~Vega and F. Woynarovich, J. Phys. A {\bf 25},  4499  (1992).

\bibitem{alma:93}
S.~R. Aladim and M.~J. Martins, J. Phys. A {\bf 26},  L529  (1993).

\bibitem{vegx:94}
H.~J. de~Vega, L. Mezincescu, and R.~I. Nepomechie, Phys. Rev. B {\bf 49},
  13223  (1994).

\bibitem{anjo:84}
N. Andrei and H. Johannesson, Phys. Lett. A {\bf 100},  108  (1984).

\end{thebibliography}
\end{document}